\documentclass[sigconf]{acmart}

\AtBeginDocument{%
  }

\usepackage{dirtytalk}

\copyrightyear{2026}
\acmYear{2026}
\setcopyright{cc}
\setcctype{by}
\acmConference[CUI '26]{ACM Conversational User Interfaces 2026}{July 21--24, 2026}{Bremen, Germany}
\acmBooktitle{ACM Conversational User Interfaces 2026 (CUI '26), July 21--24, 2026, Bremen, Germany}
\acmDOI{10.1145/3816046.3816271}
\acmISBN{979-8-4007-2741-2/2026/07}

\begin{document}

\title{Navigating Transitions: Envisioning Conversational User Interfaces to Support International Students}

\author{Yuhui Xu}
\orcid{0000-0002-3136-2149}
\affiliation{%
  \institution{Eindhoven University of Technology}
  \city{Eindhoven}
  \country{Netherlands}
}
\email{y.xu1@tue.nl}

\author{Isabel Blijenburg}
\orcid{0009-0005-4609-0560}
\affiliation{%
  \institution{Eindhoven University of Technology}
  \city{Eindhoven}
  \country{Netherlands}
}
\email{isabelblijenburg@gmail.com}

\author{Bhakti Moghe}
\orcid{0009-0003-5054-2086}
\affiliation{%
  \institution{Eindhoven University of Technology}
  \city{Eindhoven}
  \country{Netherlands}
}
\email{b.moghe@tue.nl}

\author{Maarten Houben}
\orcid{0000-0002-6571-1925}
\affiliation{%
  \institution{Eindhoven University of Technology}
  \city{Eindhoven}
  \country{Netherlands}
}
\email{m.houben1@tue.nl}

\author{Daniel Tetteroo}
\orcid{0000-0002-2295-5022}
\affiliation{%
  \institution{Eindhoven University of Technology}
  \city{Eindhoven}
  \country{Netherlands}
}
\email{d.tetteroo@tue.nl}

\author{Wijnand IJsselsteijn}
\orcid{0000-0001-6856-9269}
\affiliation{%
  \institution{Eindhoven University of Technology}
  \city{Eindhoven}
  \country{Netherlands}
}
\email{w.a.ijsselsteijn@tue.nl}

\author{Minha Lee}
\orcid{0000-0002-7990-9035}
\affiliation{%
  \institution{Eindhoven University of Technology}
  \city{Eindhoven}
  \country{Netherlands}
}
\email{m.lee@tue.nl}

\renewcommand{\shortauthors}{Xu et al.}

\begin{abstract}
International students face struggles when adapting to the host country. They are more susceptible to mental health problems than domestic students. While Conversational User Interfaces (CUIs) are increasingly researched and implemented, research on how they may help international university students is still scarce. Thus, we conducted participatory design workshops with international students who shared their perspectives and struggles of studying abroad, in which they also envisioned CUIs as aids to support their transitions. Participants proposed features of a CUI to address uncertainty, loneliness, and misunderstandings of cultural differences. Our paper reveals international students' needs and provides design implications for CUIs to support their well-being.
\end{abstract}

\ccsdesc[500]{Human-centered computing~Empirical studies in collaborative and social computing}

\keywords{international students, conversational user interface, life transition, acculturation, participatory design}

\begin{teaserfigure}
\centering
  \includegraphics[width=1.0\textwidth]{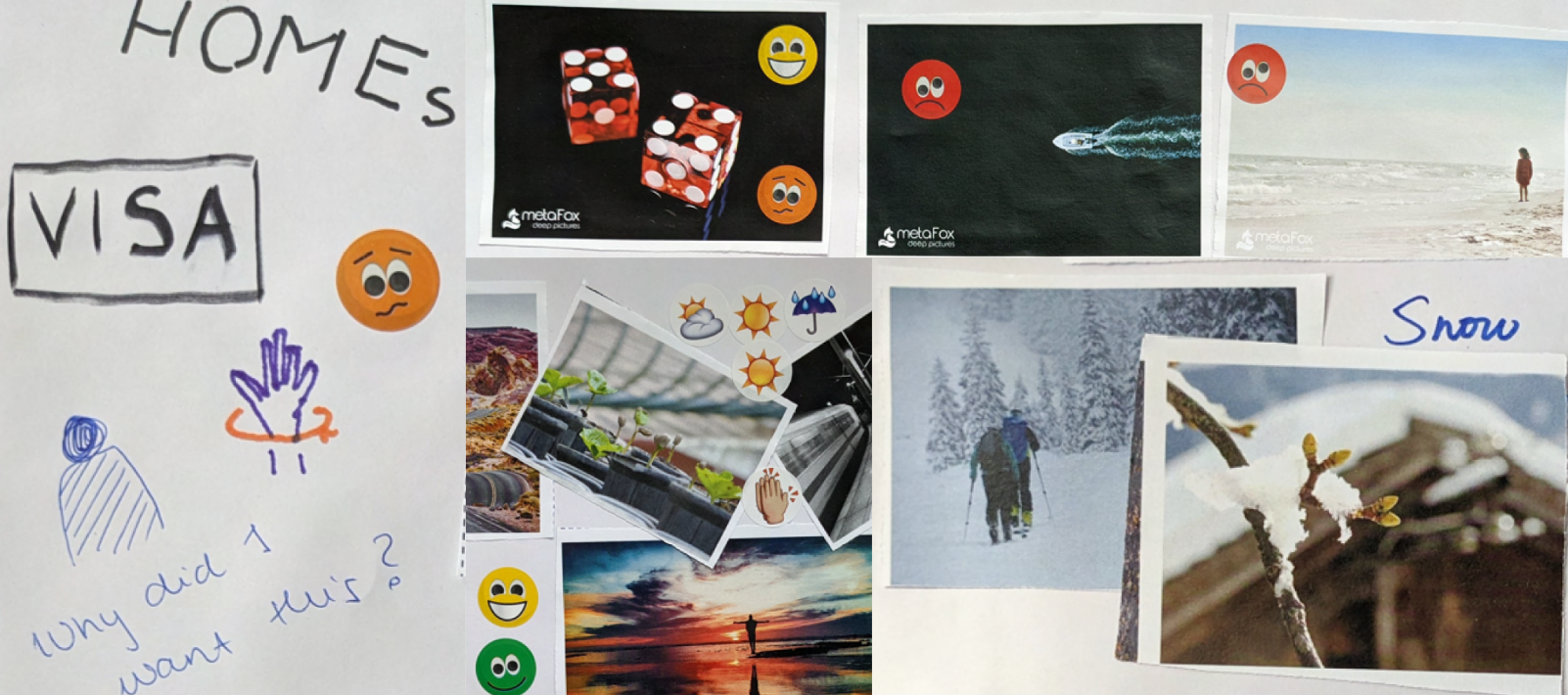}
  \caption{A collection of photos and collages made by participants in the workshops.}
    \label{fig:collection}
    \Description{A collection of photos and collages made by participants.}
\end{teaserfigure}

\maketitle
 
\section{Introduction}

Starting as a university student in a foreign country is a challenging and typical transition period \citep{van2019rites, Kahu2020UsingLearning}. International students face additional cultural difficulties \citep{Ozer2015PredictorsDenmark} and unique sources of stress such as language barriers, academic performances, cultural gaps, and financial difficulties \cite{mori2000addressing}. Furthermore, moving to a different country suddenly separates international students from their established support system \citep{pedersen1991counseling}. These struggles may lead to acculturative stress \citep{Berry2006AcculturativeStress} and a high risk of developing mental health problems, such as distress, anxiety, and depression \citep{Forbes-Mewett2016InternationalHealth}. Research reported how international students used different coping mechanisms to lower acculturative stress, such as interacting with domestic students \citep{Thomas2018LeveragingEngagement} or receiving support from their parents \citep{Crockett2007AcculturativeStudents.}. However, international students tended to be hesitant to get professional support mainly due to different cultural constructions or cultural expressions of personal distress \citep{Forbes-Mewett2016InternationalHealth, Han2013ReportUniversity, Zhou2021PrevalenceU.S.}. Meanwhile, despite the growing attention from HCI and CSCW on students' mental health \cite{mckay2010mental, lattie2020designing, yoo2019designing}, how international students manage the additional stressors remains underexplored. Therefore, there is a need to explore ways to provide alternative and accessible forms of support for dealing with struggles and vulnerabilities that are specifically targeted at international students. 

There are many forms of technologies that provide support to individuals' well-being, such as CUIs. Conversational User Interfaces (CUIs) are an emerging research area to support well-being \citep{Denecke2021ATest, lee2019caring, Ahmed2021AFeatures}. Research has also explored the role of CUIs in supporting the general student population \citep{Dibitonto2018ChatbotLife, Heo2019CiSA:Academics}. However, given the unique struggles and vulnerabilities of international students, CUIs for supporting this specific group's transitions are not yet well-studied, thus calling for more insights on how CUIs can be designed for this purpose. 

In this study, we sought design opportunities in CUIs to provide support, based on understanding the international students' experiences and struggles during transitions. We adopted a participatory design approach by conducting two workshops with visual collages (i.e., craft materials to express people's experiences visually) \citep{Sanders2020ConvivialDesign,pain2012literature,glaw2017visual,pauwels2015participatory} to actively involve international students as experts in their own experiences. Our findings report design implications of CUIs in supporting international students, based on understanding their experiences from the visual collages they created.

\section{Related Work}

\subsection{Transition and Acculturative Stress}

International students starting their college life are a typical group experiencing life transition \citep{van2019rites}. They experience separation from established social networks (e.g., family and friends) \citep{weiss1975loneliness}, past and new identities in flux \citep{constantine2005qualitative, leong2015coming}, and incorporation of new identities and social networks \citep{marginson2014student}. In such transitional processes, they may experience acculturative stress as a unique stressor than local students \citep{mori2000addressing}. Acculturation is the process of cultural and psychological change when different cultures come into contact \citep{Sam2010Acculturation:Meet}. One of the possible outcomes of acculturation is acculturative stress, describing problematic but controllable acculturative experiences  \citep{Berry2006AcculturativeStress} and a reduction in the physical and mental health of individuals facing acculturation difficulties \citep{Berry1987ComparativeStress}. Acculturative stress can negatively affect mental health \citep{Berry1987ComparativeStress}, lead to more academic stress \citep{Misra2003RelationshipsStates., Rienties2012UnderstandingIntegration}, and be a predictor for depression \citep{Zhang2012AnStudents, Smiljanic2017TheSymptoms, Constantine2004Self-ConcealmentStudents.}. One identified predictor of acculturative stress is language fluency: higher fluency and higher frequency of use of the host language could lead to lower levels of acculturative stress \citep{Yeh2003InternationalStress, Poyrazli2004SocialStudents, Sumer2008PredictorsStudents} and may improve one's sense of belonging \citep{Beukema2020SocialStudents}. These acculturation situations that are difficult to understand without similar cultural experiences may lead to international students keeping their struggles to themselves without disclosing them, even to close relationships \citep{constantine2005qualitative}. However, their transitional and acculturative experiences drew little attention from the HCI and CSCW communities \citep{binsahl2015identity, sabie2022decade} to research on technological supportive approaches. Thus, we tried to address this gap through this study to support this group's unique needs.

\subsection{Supporting International Students}

Social support includes emotional, institutional, and instrumental support \citep{Franco2019AcculturativeStudents} to develop coping strategies for complex and stressful experiences like moving abroad \citep{Franco2019AcculturativeStudents}. However, studying abroad separates international students from their previously established social support systems \citep{pedersen1991counseling}. Meanwhile, they often experience high academic pressure and therefore prioritize academics over socializing \citep{Curtin2013FosteringStudents}, making it more challenging to develop a new support system \citep{Brunsting2021SourcesStudents}. International students can receive instrumental support from host country residents, such as helping out with rules and regulations \citep{Ong2005TheSojourners}, while emotional support can only be accessed from distant family and friends and fellow international students \citep{Ong2005TheSojourners, Chavajay2013PerceivedUniversity}. 

There are existing ways of supporting international students, but they are not well-accepted by international students due to various reasons, such as cultural differences. International students tend to under-utilize university counseling services even with mental health issues, often due to insufficient information about the services (other challenges and problems) \citep{Russell2008InternationalServices, Zhou2021PrevalenceU.S.}. Additionally, some international students do not want to use these counseling services because of the barriers of language and understanding different cultures \citep{Ang2008OutServices, Mukminin2019AcculturativeEducation, Han2013ReportUniversity, Russell2008InternationalServices}. Another strategy is information-seeking while adapting to studying abroad \citep{natalie2018understanding}, e.g., searching online for information about the host country \citep{Khawaja2011UnderstandingApproach}. Also, joining a student club or association can also help with the acculturative transition of international students \citep{Menzies2014InternationalFriends, Sharma2016RoleStudents}. However, international and domestic students can show little engagement with each other \citep{Arkoudis2013FindingEducation, Thomas2018LeveragingEngagement} due to, e.g., a preference to socialize with students with a similar cultural background and language \citep{Eisenchlas2007DevelopingInteraction}.  

\subsection{Conversational User Interfaces for Supporting Transitions}

Many of the digital tools are developed based on opinions from mental health experts (e.g., psychologists) instead of the end-users. Moreover, research specifically focused on international students as end-users is lacking. With the increasing demand for technology-supported well-being, CUIs have been focused on various aspects. From an information retrieval perspective, CUIs have been used to provide access to local information for people coming from other countries for jobs or education \citep{abbas2023chatbots}. Also, CUIs have been studied as alternatives or supplements to face-to-face services, e.g., chatbots that deliver conversational cognitive behavioral therapy \citep{Fitzpatrick2017DeliveringTrial}. Another approach is giving people a caregiver role towards a CUI to help them care for themselves \citep{Lee2019CaringSelf-compassion}. 

There is a research gap on how to support international students in dealing with struggles and vulnerabilities, while we saw CUIs' potential to provide help. CUIs can help mediate students' needs and available services \citep{lee2017bots}. CUIs relieve certain barriers in mental health care, such as the stigma surrounding psychological disorders in some cultures \citep{Miner2016ConversationalHealth}. 
A CUI can increase the accessibility of relevant information for international students \citep{Heo2019CiSA:Academics} or provide notifications about courses, university events, and academic information \citep{Dibitonto2018ChatbotLife}. Additionally, CUIs can provide emotional support for one's life transformation and growth \citep{lopatovska2022capturing}. Therefore, this paper aims to investigate how a CUI can help students by involving them in a participatory design process.

\section{Method}

We took a participatory design approach with visual means (e.g., collages and sketches) to identify acculturation issues as well as opportunities for designing supportive CUIs. We emphasized creativity to stimulate open discussions on participants' difficult experiences by providing a safe space for self-expression through visual collages \citep{pain2012literature,glaw2017visual,pauwels2015participatory, Sanders2020ConvivialDesign}. The international students also designed chatbot features to imagine and express their ideas for supportive strategies \citep{Sanders2020ConvivialDesign}.

\begin{table*}
  \caption{Eight international students participated in two workshops}
  \label{tab: demographics}
  \begin{tabular}{ccccccc}
    \toprule
    & Workshop & Home country & Nationality & Age & Gender & Time since arrived\\
    \midrule
    SG        &1    & Turkey       & Turkish     & 25  & Woman & 3 months     \\
    GS        &1    & Turkey       & Turkish     & 25  & Woman & 3 months     \\
    NE        &1    & Iran         & Iranian     & 28  & Woman & 4 months    \\
    XZ        &1    & China        & Chinese     & 31  & Woman & 7 years   \\
    YY        &2    & China        & Chinese     & 24  & Man & 1 year    \\
    AN        &2    & Greece       & Greek       & 30  & Man &  5 years    \\
    DB        &2    & Hungary      & Hungarian   & 23  & Woman & 1 year    \\
    DM        &2    & UAE          & Portuguese  & 19  & Woman & 1.5 year    \\
    \bottomrule
  \end{tabular}
\end{table*}

\subsection{Participants}

We reached out to international students of Eindhoven University of Technology via student associations and email invitations, and collected their demographic data. We recruited 2 Turkish, 2 Chinese, 1 Iranian, 1 Greek, 1 Hungarian, and 1 Portuguese participants, with their demographics illustrated in Table \ref{tab: demographics}. Participants each received a compensation of 20 Euros. Pseudonyms are used for the participants to ensure confidentiality. 

\subsection{Workshop}

We hosted two separate two-hour workshops with different groups of four participants in each, to explore design opportunities for CUIs to provide support for international students. At the start of the workshop, the participants were given information about the study and were asked for informed consent. In the beginning, the participants were provided with craft materials to make a collage that showed their most important memories during their stay here. Secondly, the created collages sparked deeper conversations where participants described to the group what they made and what it represented. The participants responded to each other's stories and figured out together which acculturation issues they had experienced, which were made into an overview. Thirdly, the participants were encouraged to sketch conversational interfaces, text-based conversations, other functionalities of the CUI, and design opportunities for the aforementioned experienced problems. Finally, the participants added comments and stickers to each other's prototypes and discussed their sketches openly.  

\subsection{Data Analysis}

The audio recordings of the workshops were transcribed. We, as a group of researchers with mixed cultural backgrounds, conducted a preliminary thematic analysis on the audio transcripts and notes using an inductive approach \citep{clarke2015thematic}. The second author collected all the data and created the initial set of codes, which were then iterated with the last author into a first set of themes that informed the next research stage. Afterward, all authors looked into the data to arrive at our final themes. Reflecting on our positionality, the author group consists of four international PhD students (two current, two past) and three domestic scholars. 

\section{Results}

\begin{figure}
  \centering
  \includegraphics[width=1\linewidth]{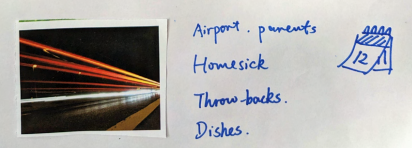}
  \caption{YY's illustration represented his taxi trip to the airport when he was experiencing uncertainty.}
    \label{fig:YY_airport}
    \Description{Left: A photo picked by participant YY to represent his taxi trip to the airport. Right: Participant YY's handwriting that says "Airport presents homesick. Throw-backs. Dishes" and a calendar icon.}
\end{figure}

The collages expressed the participants' acculturation experiences and struggles in transition (e.g., Fig. \ref{fig:YY_airport}) and showed differences in how each student uniquely visualized their memories, which represented their well-being and self-identity through metaphors. They also sketched a potential CUI's features and offered insights into strategies to address their struggles.

\subsection{Uncertainty and Doubt}

\begin{figure}
  \centering
  \includegraphics[width=1\linewidth]{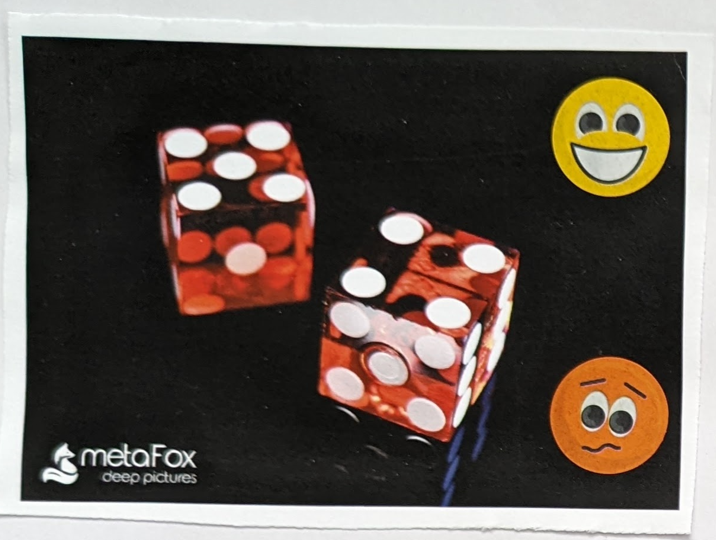}
  \caption{XZ used this image and emoticon stickers as a metaphor for uncertainty.}
  \label{fig:XZ_dice}
  \Description{A photo of two dice with a happy face emoji in the top-right corner and a sad face emoji in the bottom-right corner, made by participant XZ as a metaphor for uncertainty.}
\end{figure}

Moving abroad fundamentally altered the living and cultural environments of the international student participants. Familiar routines, systems, and cultures of everyday life were replaced by new ways of functioning, while the participants initially lacked knowledge of these unfamiliar norms and practices that they needed to re-learn and adapt to. These disruptions became a source of uncertainty and doubt, as participants navigated between the known and the unknown. Yet, embedded within these changes were both positive and negative possibilities, revealing how uncertainty can simultaneously open pathways for growth and pose challenges to adaptation. Participants experienced many positive aspects of moving abroad, e.g., \textit{\say{I'm really brave that I made this decision to move to another country}} (NE) and \textit{\say{When I realized that we're all superheroes for that [doing stuff in a different language] alone, that was a big moment for me}} (DB). However, moving to the Netherlands was met with a lot of uncertainty. As visualized by YY in Figure \ref{fig:YY_airport}: \textit{\say{When I decided to study abroad, I didn't hesitate, but when I was in the taxi to the airport, that's when I [thought] I don't want to do that}} (YY). XZ described this feeling through the dice image as illustrated in Figure \ref{fig:XZ_dice}: \textit{\say{When I first came here, it was like throwing a dice, so you don't know what will happen: could be nice or could be bad, just uncertain}}. These revealed experiences implicate an opportunity for support to ease the uncertainty by providing sufficient information on lives and cultures.

\textbf{\textit{CUI for exchanging information to address uncertainty:}} Participants came up with desired features of CUIs that may help address their feelings of uncertainty and doubt. DM recommended a \textit{\say{starter package}} where international students can get answers to common questions like \textit{\say{How can I rent a bike?}} or \textit{\say{Which housing agencies should I check?}}, and also \textit{\say{How to open a bank account or find a doctor, etc}} (NE). For housing, SG suggested that international students could be matched with each other to find shared housing. A related idea is to fast-track university integration with a chatbot that serves as a study buddy or connects international students to course group members (AN). A lot of information exists online, but integration across them all is difficult, so DB mentioned a chatbot that \textit{\say{sits on top}} of various existing applications or websites could help.

\subsection{Loneliness and lack of support}

Participants' accounts revealed that navigating a new living and cultural environment often meant facing uncertainty alone. They were solely responsible for finding information and resources, yet the paths to support were often unclear. As emotional struggles accumulated without outlets for release, feelings of isolation deepened, making adaptation feel like a solitary battle against an unfamiliar world. For example, AN made all practical arrangements on himself, such as dealing with health insurance, finding housing, and a Dutch identity card: \textit{\say{Anything that has to do with legal [things] usually stresses me out. And when you do it from a different country, not exactly knowing the system here, every little detail is so magnified}} (AN). Experiencing such problems can also lead to a negative mood according to SG: \textit{\say{During those days, having no one to consult about what to do, you feel even more lonely.}} The lack of support in arranging such matters often leads to feeling overwhelmed and alone: \textit{\say{Because you are here all alone everything is overwhelming. You are feeling alone in an unknown territory [...] What do I do? I have no idea [...] Everything is just too much}} (XZ). GS also used a road image to describe her situation of \textit{\say{feeling very alone}}. Participants could cope with experiencing a new journey but had vulnerable moments. NE put together different emoticon stickers to show her ups and downs: \textit{\say{I experienced a lot of different emotions. I was happy. And in the evening I suddenly felt sad and I cried a lot. I missed my family}} (NE). These illustrate how uncertainty was not only cognitive but deeply emotional, arising from both practical struggles and the absence of relational or institutional support, and thus making adaptation a process of constant negotiation between independence and vulnerability.

\begin{figure}
  \centering
  \includegraphics[width=1\linewidth]{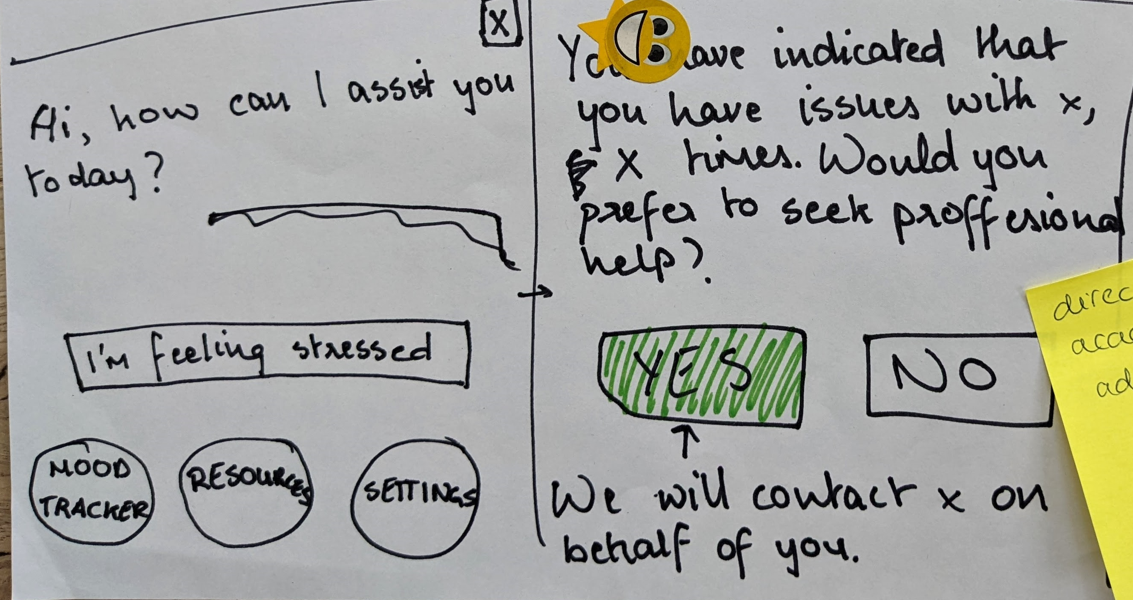}
  \caption{DM's sketch of a chatbot that provides several mental health resources.}
  \label{fig:DM_mental_health}
  \Description{A sketch made by participant DM that shows a rough chatbot interface that provides several mental health resources.}
\end{figure}

\textbf{\textit{CUI for community-driven approach to address loneliness:}} Support from communities is critical to help international students cope with loneliness. Participants envisioned conversational interfaces that not only provided information but also facilitated social connection and emotional reassurance. For instance, XZ suggested pairing newly arrived international students with domestic or senior international students, or connecting them to clubs and associations. Building on this, SG imagined a chatbot that could answer questions such as \textit{\say{How can I make new friends, chatbot?}} by offering actionable links in reply: \say{Join this WhatsApp group we created for people who asked the same question}. AN and DM emphasized the importance of tracking users' emotional state over time, so that the CUI could recommend personalized activities or resources when signs of stress appear. When students are not doing well, e.g., experiencing stress, DM recommended that various resources, such as online videos, can benefit international students (see Figure \ref{fig:DM_mental_health}). Moreover, they proposed that chatbots could normalize emotional struggles by sharing relatable peer experiences, e.g., \textit{\say{James from Applied Physics faced the same issue, and this is what helped him.}} These insights point to a community-driven CUI design that goes beyond one-to-one support. Such systems could function as social mediators, connecting users to both informational and emotional resources while cultivating a shared sense of belonging among international students.

\subsection{Misunderstanding of cultural differences}

Access to social support did not necessarily translate into a sense of connection. The cultural environment itself introduced subtle yet pervasive differences that complicated how international students related to others. These differences were not always noticeable, but they quietly influenced emotions and everyday interactions. Over time, such experiences heightened participants' awareness of their cultural identities and the potential for misunderstanding or bias in cross-cultural communications. Reflecting on living in the Netherlands, participants noted that there is open-minded curiosity and ignorance: \textit{\say{My friend from Mexico got a question if they have ATMs in Mexico. Or Dutch will talk about Africa like it's their country}} (AN). Participants were aware that their own ethnicity and background impact their likelihood of experiencing discrimination, e.g., \textit{\say{I'm a white girl, and that's why I don't experience the negative side}} (DB). At the same time, one can be sensitive to discrimination regardless of one's ethnicity or background: \textit{\say{I'm also white and a male, so it's not that I'm the target of much hate}} (AN). Given the awareness of cultural differences that may lead to cultural misunderstandings, many participants came up with coping strategies. For example, XZ shared the thought of self-initiated introduction of each other's culture: \textit{\say{It would be nice if Dutch students and international students could have some kind of session to introduce each other's culture. So you know what to expect from another person.}} Accordingly, external forces are often needed to help identify and mediate cultural differences, as well as to provide information and access to relevant support systems.

\textbf{\textit{CUI for understanding and mitigating cultural differences:}} CUIs can help international students navigate and interpret cross-cultural misunderstandings that may otherwise lead to discomfort or harm. Participants emphasized that when facing ambiguous or potentially discriminatory situations, students often attributed such experiences to cultural misunderstanding rather than intentional bias. This self-directed interpretation reflected their desire to integrate and maintain harmony, even at the cost of overlooking negative experiences. For instance, NE envisioned a chatbot that could assist in assessing these situations: \textit{\say{You can search or ask the chatbot about this behavior I experienced. What was wrong? Did I make a mistake? Or is it strange?}} (NE). Such a system could act as a reflective mediator to help students make sense of cultural nuances, validate their perceptions, and offer guidance on how to respond constructively. These insights suggest an opportunity to design CUIs that scaffold cultural interpretation, helping users contextualize behaviors, recognize intents, and access supportive resources without worry of over-reacting or misunderstanding.

\section{Discussion}

International students relocating to a new environment face difficulties in adaptation and may experience acculturative stress. Given that CUIs for providing corresponding support are not yet well-studied, more CUI design insights are needed to support international students' acculturation and transitions. This study contributes by revealing international students' need for easier access to local information to address the feelings of uncertainty and doubt. This is in line with the previous work from the HCI community on CUI designs for college students' mental health \citep{crasto2021carebot, dhanasekar2021chatbot, de2020investigating} and CUIs for facilitating information retrieval and efficient communication \citep{fabian2023adapting, di2018chatbots}. Additionally, the conversational nature of CUIs provides space for personalized dialogues \citep{chen2020creating} that could be tailored for international students to allow for empathetic interactions to address loneliness and lack of support. 

With a focus on international students, our paper contributes to an emerging research line in CSCW and HCI on CUIs for supporting the well-being of vulnerable groups (e.g., international students). Our participatory design approach offered qualitative insights into international students' experiences and strategies for CUIs to provide support. We further discuss how international students' acculturation process can be aided by CUIs in 1) mediating access to official institutions and practicalities, 2) bridging communities of students for peer support, and 3) exploring cultural differences.

Our findings highlight several opportunities for designing CUIs to support international students' transition and adaptation. First, CUIs could act as a starter package, providing practical information upon arrival and answers to common questions. Second, they could serve as a personalized information aggregator, helping students integrate guidance from multiple websites and administrative sources. Third, CUIs could facilitate social connection, linking newcomers to peers, communities, or clubs to foster a sense of belonging. Finally, CUIs could mediate cultural and social ambiguities and differences, supporting students in interpreting norms and navigating unfamiliar social practices. Together, these functionalities suggest a versatile, student-centered conversational system that addresses both informational and socio-emotional needs during the transition to a new environment.

\subsection{Mediating Access to Official Institutions and Practicalities}

Any CUI that supports integration and acculturation should be treated as a node in larger societal or governmental networks \citep{abbas2023chatbots}. An opportunity for chatbots is to support people in gaining access to practical information in new contexts, which is difficult to find for newcomers. Our findings reveal that CUIs can support international students in times of drastic changes in living arrangements by providing easy access to practical information, which can reduce stress and worries caused by unclear regulations, missing information, or a lack of assistance from official institutions \citep{Ozer2015PredictorsDenmark, Poyrazli2004SocialStudents, Yeh2003InternationalStress}. Many participants struggled to navigate across foreign university and government websites; a CUI could provide a low barrier for international students to enter a new cultural world that they do not fully understand. CUIs can overcome potential language barriers and support official instances to cater to people who cannot speak the official language. For example, a CUI that shows images describing the topic might improve language skills, which are related to lower acculturative stress \citep{Poyrazli2004SocialStudents, Sumer2008PredictorsStudents, Yeh2003InternationalStress}. As such, CUIs are a suitable solution to support access for people disadvantaged by a lack of language skills or social networks to official institutions, such as universities, government institutions, or housing or insurance agencies.

\subsection{Bridging Communities of Students for Peer-Support}

We demonstrated that international students saw value in CUIs that facilitate experience sharing and meetups to address loneliness or a lacking sense of social belonging. Similarly, access to online peer support was crucial for technologies that aimed to support young adults' mental health \citep{dauden2022exploring}. Our participants experienced difficulties balancing study life and leisure activities, which are crucial for meeting new people and building a social network. These findings expand existing research demonstrating how a CUI supports the general student population in coping with academic stress \citep{de2020investigating} or procrastination \citep{pereira2021struggling} by highlighting the specific social needs of international students. For example, the participants expressed the challenges of being separated from friends and family while not knowing how and where to meet new people, as international and domestic students tend to have little interaction with each other \citep{Arkoudis2013FindingEducation, Thomas2018LeveragingEngagement}. Importantly, only routing international students to international associations may not be inclusive (SG), so a CUI can better suggest diverse clubs and associations that will bring local and international students together. Furthermore, novel ways of matching students based on shared needs, interests, and preferences by CUIs can be explored in future research. 

\subsection{Exploring Cultural Differences}

International students tended to focus on positive aspects of intercultural exchange, e.g., a CUI suggesting fusion cuisine (DB), but they did experience challenges related to cultural differences, such as misunderstanding of behavior or cultural norms of their host country, which is often faced by people with a different background \citep{Poyrazli2004SocialStudents, Yeh2003InternationalStress}. In this context, CUIs can provide a safe and accessible platform for questions related to cultural practices and habits that international students are hesitant to ask domestic students. Participants saw the benefit of CUIs that facilitate cultural exchange and social integration by providing emotional assurance. However, we note that international students may be more prone to being accepted into a new culture, often at the cost of shedding their original cultural identity \citep{teske1974acculturation}. University students are young adults, looking for group affiliation and approval, yet they also receive specific comments about not speaking Dutch, resulting in additional stress. This complicates the picture of what constitutes healthy acculturation as international students as young adults can take on new cultural identities without needing to feel \say{inferior} and deserve special attention. 

\subsection{Risks and Limitations of CUIs for Transitional Support}

While CUIs hold potential in supporting international students' transitions, their deployment should be approached with caution. Our findings highlighted participants' experiences of uncertainty, loneliness and lack of support, and cultural misunderstanding, which, however, are contexts that could make users particularly vulnerable to over-reliance on CUIs. Recent HCI scholarship warns that empathic or supportive chatbots may create emotional dependency by simulating empathy without true understanding \citep{kurian2025ai}. When experiencing doubt and loneliness, users may seek comfort from CUIs as a substitute for human interaction, potentially reducing motivation to build genuine social connections \citep{skjuve_my_2021}. Moreover, the cultural framing of empathy within CUIs needs careful consideration. As our participants' reflections on misunderstanding and stereotypes suggest, a chatbot's \say{empathic} responses may reflect cultural biases embedded in its training data, reinforcing rather than alleviating miscommunication \citep{basharat_ai-enabled_2024}. Additionally, CUIs also have the risk of misinformation by providing incomplete or incorrect guidance when international students face complex administrative or cultural questions, leading to potentially harmful consequences \citep{meyrowitsch_frontiers_2023}. Therefore, while CUIs offer accessible and responsive interaction, their design for transitional support must critically consider not only efficacy but also ethical and emotional safety, ensuring that such systems complement rather than replace human care and institutional support.

\section{Conclusion}

We brought together research agendas on CUIs to support international students' struggles in transition. By building on prior HCI and CSCW research on college students' challenges, we highlighted that international students are a particular subgroup that has been less researched. We thus explored international students' college experience through participatory design workshops, where participants envisioned how CUIs can support them in dealing with struggles in transition. Qualitative insights revealed struggles in transitions and opportunities for CUIs to provide support. While CUIs may support in various aspects, they also risk simplifying complex transitional identities into scripted interactions, thus calling for more careful and nuanced design. Future research can also address how bi-directional cultural exchange can be facilitated through CUIs and how institutional support should be fostered.

\bibliographystyle{ACM-Reference-Format}
\bibliography{CUI25_template}

\end{document}